\documentclass[aps,prd,reprint,amssymb]{revtex4-1}
\usepackage{amsmath}
\usepackage{graphicx}
\usepackage{dcolumn}
\newcommand{\RomanNumeralCaps}[1]
    {\MakeUppercase{\romannumeral #1}}
\usepackage{bm}

\begin{document}

\title{Grassmannian sigma model on a finite interval}

\author{D. Pavshinkin}
 \email{dmitriy.pavshinkin@phystech.edu}
\affiliation{
 Institute of Theoretical and Experimental Physics, Moscow 117218, Russia\\Moscow Institute of Physics and Technology, Dolgoprudny 141700, Russia
}

\date{1 September 2017}

\begin{abstract}
We  discuss the two-dimensional Grassmannian sigma model {\small$\mathbb{G}_{N, M}$} on a finite interval {\small$L$}. The different 
boundary conditions which allow one to obtain analytical solutions by the saddle-point method in the large {\small$N$} limit
are considered. The  nontrivial phase structure of the model on the interval similar to the {\small$\mathbb{C}P(N)$} model 
is found.
\end{abstract}

\maketitle


\section*{\RomanNumeralCaps{1}. Introduction} 
 Two-dimensional sigma models are of great interest because of their similarity with the four-dimensional Yang-Mills theory and serve as a good platform for studying nonperturbative effects \cite{polyakov, nsvz}. For the first time they [{\small$\mathbb{O}(N)$} and {\small$\mathbb{C}P(N)$}] were solved in \cite {polyakov, dadda, witten} in the  large {\small$N$} limit \cite{moshe}.
The {\small$\mathbb{C}P(N)$} and Grassmannian sigma models also emerge as the worldsheet theories on the single and multiple non-Abelian strings
in QCD-like nonsupersymmetric theories \cite{gsy,sakai}( see \cite{sy,tong} for review). Therefore the corresponding models on the interval 
correspond to the non-Abelian strings stretched between the boundary branes representing the domain walls. The {\small$\mathbb{C}P(N)$} model
on the interval has been considered in \cite{milekhin1,bolognesi,milekhin2}. It was found that the solution of the model strongly
depends on the boundary conditions and for the particular choice of the boundary conditions can undergo the phase transition
at some value of the length of the interval.
 
 In this paper we consider field theory on the complex Grassmannian manifold \cite{perelomov}
{\small $$\mathbb{G}_{N,M}=\frac{SU(N)}{SU(N-M)\times U(M)},\;\;\;M<N.$$
}
 The fields $\varphi(x_0,x_1)$ are a complex matrix {\small$N\times M$} satisfying the additional conditions $\varphi^\dagger \varphi=I_M$ (unit matrix {\small$M\times M$}), hence it is a set of {\small$N$}-dimensional vectors orthogonal to each other.
 We will get effective action and will  study the large {\small$N$} solution of the saddle-point equation with various types of boundary conditions:  Dirichlet-Dirichlet(D-D), Neumann-Neumann(N-N), and  mixed Dirichlet-Neumann(D-N). Similar studies were carried out in  \cite{milekhin2, bolognesi} for {\small$\mathbb{C}P(N)$} model {\small$\big(\mathbb{C}P(N)=\mathbb{G}_{N,1}\big)$}. In \cite{milekhin2} boundary conditions admitting an analytical solution have been found and which yields the existence of two phases: Higgs phase with broken {\small$U(1)$} gauge symmetry and Coulomb phase with unbroken {\small$U(1)$} symmetry. We will  show that the similar result is true for {\small$\mathbb{G}_{N,M}$} as well.\\

\section*{\RomanNumeralCaps{2}. Effective action}
Let us remind the reader of the Lagrangian of the model. The unique {\small$SU(N)$}-invariant metric on the {\small$\mathbb{G}_{N,M}$} is  $ds^2={\small Tr(d\varphi d\varphi)}$; hence the Lagrangian reads as
{\small $$L=Tr\big[N/g^2(D_\mu \varphi)^\dagger D_\mu \varphi+\lambda(I_M-\varphi^\dagger\varphi)\big],\eqno(1)$$ }
  where $\lambda$ is Lagrange multiplier,
 $g$ is the coupling constant, {\small$D_\mu=\partial_\mu-iA_\mu$} and {\small$x\in[0,L],\;t\in(-\infty,+\infty)$}.

 We consider the partition function in the Euclidean space:
{\small$$
Z = \int DAD\lambda D\varphi^\dagger D\varphi$$
$$\times\exp\bigg[-\int d^2x Tr\Big(N/g^2(D_\mu \varphi)^\dagger D_\mu \varphi+\lambda(\varphi^\dagger\varphi-I_M)\Big)\bigg]$$
$$=\int DAD\lambda D\varphi_{ij}D\varphi_{ij}^*$$
$$\times\exp\bigg[-\int d^2x\Big(N/g^2(\partial_\mu-iA_\mu)\varphi^*_{ij}(\partial_\mu+iA_\mu)\varphi_{ij}+$$
$$+\lambda(\varphi^*_{ij}\varphi_{ij}-M)\Big)\bigg]\;\eqno(2)
$$
}
 Integrating by parts and taking into account the relation $\varphi^\dagger \varphi=I_M$, we obtain
{\small$$ \int d^2x(D_\mu \varphi_{ij})^* D_\mu \varphi_{ij}=-\int  d^2x\varphi_{ij}^*D_\mu ^2 \varphi_{ij}
 $$}

 and 
 {\small$$Z = \int DAD\lambda DRe\varphi_{ij}DIm\varphi_{ij} \exp\bigg[-$$  
 $$-\int d^2x\Big(Re\varphi_{ij}\Big(-N/g^2(\partial_\mu+iA_\mu)^2+\lambda \Big)Re\varphi_{ij}+$$
 $$+Im\varphi_{ij}\Big(-N/g^2(\partial_\mu+iA_\mu)^2+\lambda \Big)Im\varphi_{ij}-\lambda M\Big)\bigg].\eqno(3)$$  
 }
 In more detail, the quantity in the exponent is 
 {\small$$ \sum_{i,j}\bigg(-\int d^2x Re\varphi_{ij}\Big(-\frac{N}{g^2}D_\mu  ^2+\lambda \Big)Re\varphi_{ij}\bigg)+$$ $$+ \sum_{i,j}\bigg(-\int d^2x Im\varphi_{ij}\Big(-\frac{N}{g^2}D_\mu ^2+\lambda \Big)Im\varphi_{ij}\bigg)+\int d^2x\lambda M.$$
 }
 This integral is Gaussian, so we can separate $\varphi_{11}=\sigma$ and easily integrate out the remaining {\small$NM-1$} fields. Then we obtain an action depending only on $\lambda$ and $\sigma$.
 
 Taking into account that
 {\small
 $$ \int DRe\varphi_{ij} \exp\bigg[\sum_{i,j}\Big(-\int d^2xRe\varphi_{ij}\Big(-\frac{N}{g^2}D_\mu ^2+\lambda \Big)Re\varphi_{ij}\Big)\bigg]\sim$$ $$\sim\Big[det\Big(-N/g^2D_\mu ^2+\lambda \Big)\Big]^{-(NM-1)/2} $$
 }
 we get
 {\small
 $$\widetilde{Z}=\int DAD\lambda D\sigma D\sigma^* \exp\bigg[-(NM-1)Tr \ln\Big(-D_\mu ^2+\frac{g^2}{N}\lambda \Big)-$$
 $$-\int d^2x\Big((D_\mu\sigma)^2+\lambda \frac{g^2}{N}\sigma \sigma^* -\lambda M\Big)\bigg].$$
 }
 In the last equation the fields were  rescaled.
 Redesignating $\lambda$ and introducing the value $r= { MN / g ^ 2}$, we obtain an expression for the effective action:
{\small $$S_{eff}=(NM-1)Tr\ln\big(-D_\mu ^2 +\lambda \big)+$$
 $$+\int d^2x\Big((D_\mu\sigma)^2+\lambda(|\sigma|^2 -r)\Big).\eqno(4)$$
}\section*{\RomanNumeralCaps{3}. Saddle-point equations}
 We will calculate the path integral by the saddle-point method in the large {\small$N$} limit. For this we are looking for saddle points with respect to $\lambda$ and $\sigma$ of the action function. With the chosen boundary conditions, translational symmetry with respect to $x$ is broken, but it is conserved with respect to $t$. Therefore we consider $\lambda=\lambda (x),\;\sigma=\sigma (x).$ Let's set the gauge  {\small$A_t=0$} and consider the eigenvalues and eigenfunctions of the operator:
{\small $$\Big(-D_x^2+\lambda(x)\Big)\psi_n(x)=\mu_n\psi_n(x).\eqno(5)$$
 }
 For our boundary conditions,
 {\small$\psi_n(x)\sim \sin(E_nx+b_n)$}
 and {\small$\mu_n=E_n ^2.$}
 Thus, using the relation {\small$Tr\:ln=ln\:det$} we can represent effective action in the form:
 {\small$$S_{eff}=(NM-1)\sum_{n}E_n+\int d^2x\Big((D_\mu\sigma)^2+\lambda(|\sigma|^2 -r)\Big).\eqno(6)$$
 }
Let us vary effective action with respect to $\lambda$ and use the first-order correction in perturbation theory, then we get the first-saddle point equation: 
{\small $$\frac{NM-1}{g^2}\sum_{n}\frac{\psi_n^2(x)}{E_n}+|\sigma(x)|^2-r=0.\;\;\;\eqno(7)$$
}
 The second equation is obtained by varying action with respect to $\sigma$ which yields the equation of motion:
 {\small$$D^2_x\sigma^*-\lambda(x)\sigma^*=0.\;\;\;\eqno(8)$$}
 Further assume that {\small$A_x=0$} and  $\sigma$ is real. In this case the saddle-point equation with respect to {\small$A_x$} will be identically satisfied.
 \\
\section*{\RomanNumeralCaps{4}. D-N boundary conditions}
 Now we are considering mixed Dirichlet-Neumann boundary conditions. Let {\small$NM = 2Z +1$, $Z\in\mathbb{N}$}, and following \cite{milekhin2} suppose that on $\varphi_{11}$  the boundary conditions {\small N-N} are imposed:
 {\small$$D_x\sigma(0)=D_x\sigma(L)=0,\eqno(9)$$}
 {\small Z} fields have {\small N-D}:
 {\small$$D_x\varphi_{ij}(0)=\varphi_{ij}(L)=0,\eqno(10)$$}
 and the remaining {\small Z} fields have {\small D-N}: {\small$$\varphi_{ij}(0)=D_x\varphi_{ij}(L)=0.\eqno(11)$$} 
The eigenfunctions and eigenvalues corresponding to the boundary conditions:\\
\\
 for {\small D-N}:  {\small$$\psi_n(x)=\sqrt{\frac{2}{L}}\sin\Big[\frac{\pi x(n-1/2)}{L}\Big],\;\;E_n^2=\Big[\frac{\pi(n-1/2)}{L}\Big]^2+\lambda,$$ $$n=1,2,...,$$}
 for {\small N-D}: {\small$$\psi_n(x)=\sqrt{\frac{2}{L}}\cos\Big[\frac{\pi x(n-1/2)}{L}\Big],\;\;E_n^2=\Big[\frac{\pi(n-1/2)}{L}\Big]^2+\lambda,$$ 
 $$n=1,2,...$$}
 Substituting them into the Eq. {(7)}, we can see that dependence on $x$ disappears due to the trigonometric identity, hence $\sigma = const$.
 Then it follows from {(8)} that $\sigma\lambda=0$.

 First we set $\lambda$ not equal to zero, hence $\sigma = 0$ . In this solution the vacuum expectation valus $<\varphi_{ij}>=0$ and the {$U(1)$} symmetry is not broken. Therefore it is called ``Coulomb" phase.
 Equation {(7)} takes the following form:
 {\small$$\frac{Z}{\pi}\sum_{n=1}^{\infty}\bigg(\frac{1}{\sqrt{(n-1/2)^2+(\lambda L/\pi)^2}}\bigg)-r=0.\eqno(12) $$}\\
 In order to eliminate the divergence of the sum we represent it in the form
 {\small$$\frac{Z}{\pi}\sum_{n=1}^{\infty}\bigg(\frac{1}{\sqrt{(n-1/2)^2+(\lambda L/\pi)^2}}-\frac{1}{n}\bigg)+\frac{Z}{\pi}\sum_{n=1}^{\infty}\frac{1}{n}-r=0,\eqno(13)$$}\\
 and use {\small UV} cutoff:
{\small $$\frac{Z}{\pi}\sum_{n=1}^{\infty}\frac{\exp\Big(-\frac{n\pi}{L\Lambda_{uv}}\Big)}{n}=-\ln\bigg(1-\exp\Big(-\frac{\pi}{L\Lambda_{uv}}\Big)\bigg)\approx$$
$$\approx-\ln\Big(\pi/L\Lambda_{uv}\Big) .\eqno(14)$$}
Note that the sigma model has dynamical mass generation via dimensional transmutation: {$\Lambda^2=\Lambda^2_{uv}\exp\big(-4\pi/g^2\big).$}  This allows us to express $r$ in terms of dynamical scale $\Lambda$ and $\Lambda_{uv}$:
 $r=\frac{Z}{\pi}{\small\ln\big(\Lambda_{uv}/\Lambda\big)}.$ Thus,
 {\small$$\sum_{n=1}^{\infty}\bigg(\frac{1}{\sqrt{(n-1/2)^2+(\lambda L/\pi)^2}}-\frac{1}{n}\bigg)=\ln\bigg(\frac{\pi}{L\Lambda}\bigg).\eqno(15)$$}
 Now let us show that this phase does not exist on the entire interval.
 Indeed, the left-hand side in the power series in {\small$\lambda L$} has the form:
 {\small$$\frac{1}{\sqrt{(n-1/2)^2+(\lambda L/\pi)^2}}=\frac{1}{n-\frac{1}{2}}-4\bigg(\frac{\lambda L}{\pi}\bigg)\frac{1}{2n-1}+...\eqno(16)$$}
 Using the representation of the Riemann zeta function
 {\small$$\sum_{n=1}^{\infty}\frac{4}{(2n-1)^3}=\frac{7}{2}\zeta(3),\eqno(17)$$}
 we obtain the equation
 {\small$$\frac{7}{2}\zeta(3)\bigg(\frac{\lambda L}{\pi}\bigg)=\ln\bigg(\frac{4\lambda L}{\pi}\bigg).\eqno(18)$$}
 Therefore, the Coulomb phase exists only for {\small$L > \pi/4\Lambda$.}

 Now let $\sigma=const,\; \lambda=0$. This is corresponds to  the ``Higgs" phase, because nonzero $\sigma$ breaks {\small$U(1)$} symmetry. In this case, from the first saddle-point equation, we obtain that 
 {\small$$\frac{Z}{\pi}\sum_{n=1}^{\infty}\bigg(\frac{1}{n-\frac{1}{2}}-\frac{1}{n}\bigg)+\sigma^2=\frac{Z}{\pi}\ln\bigg(\frac{\pi}{L\Lambda}\bigg);\eqno(19)$$}
 the second Eq. (8) is satisfied.
Again using (18) we get $$\sigma^2=\frac{N}{\pi}\ln\bigg(\frac{\pi}{4L\Lambda}\bigg).\eqno(20)$$
 It can be seen that there is a solution only for   {\small$L\leq\frac{\pi}{4\Lambda}$}.
 \\
\section*{\RomanNumeralCaps{5}. D-D and N-N boundary conditions}
 Suppose now that $\varphi_{11}$ still has {\small N-N} boundary conditions, {\small$Z$} fields have {\small D-D} boundary  conditions and other {\small$Z$} field have {\small N-N} boundary conditions. In this case, we have\\
 \\
for {\small D-D}: {\small$$\psi_n(x)=\sqrt{\frac{2}{L}}\sin\Big[\frac{\pi xn}{L}\Big],\;\;E_n^2=\Big[\frac{\pi n}{L}\Big]^2+\lambda,$$ $$n=1,2,...,$$}
 and for {\small N-N}: {\small$$\psi_n(x)=\sqrt{\frac{2}{L}}\cos\Big[\frac{\pi xn}{L}\Big],\;\;E_n^2=\Big[\frac{\pi n}{L}\Big]^2+\lambda,$$ $$n=0,1,...$$}
 There is a zero mode with $n=0$ and the first saddle-point equation {(7)} has singularity  at $\lambda=0$. Hence, for given boundary  conditions the solution has only Coulomb phase with $\lambda=const\neq 0$. 
 Substituting this eigenfunctions and eigenvalues into the Eq. {(7)} and using the {\small UV} cutoff, we have:
{\small $$\frac{Z}{\pi}\sum_{n=1}^{\infty}\bigg(\frac{1}{n^2-\lambda L/\pi}-\frac{1}{n}\bigg)+\frac{Z}{\lambda L}=\frac{Z}{\pi}\ln\bigg(\frac{\pi}{\Lambda L}\bigg).\eqno(21)$$}
It is easy to see that for fixed {\small$\Lambda$} and {\small$L$}, it is always possible to find $\lambda$ that satisfies the equation, since the left-hand side, as a function of $\lambda$, takes a value on the whole {\small$\mathbb{R}$}. Thus, the only Coulomb phase exists for all {\small$L$}.
\\
\\
\section*{\RomanNumeralCaps{6}. Conclusion}
In this paper we have presented a two-dimensional  Grassmannian sigma model  {\small$\mathbb{G}_{N, M}$} on a finite interval in the large {\small$N$} limit. For special boundary conditions we have found  an  analytical solution to the  saddle-points equation. Under the { D-D} and { N-N} boundary conditions the solution has only one Coulomb phase ($\sigma=0$) for all values of the length $L$; under the mixed Dirichlet-Neumann boundary  conditions solution undergo the phase transition: there is Coulomb phase for {\small$L>\frac{\pi}{4\Lambda}$} and Higgs phase ($\sigma\neq0$) for {\small$L<\frac{\pi}{4\Lambda}$}. We note that it is also of interest to consider a phase structure of sigma model on a flag target space.\\

\section*{ACKNOWLEDGMENT}
  I would like to express my gratitude to A. S. Gorsky for suggesting this problem and for numerous useful comments.\\

\end{document}